\documentstyle[12pt,aps,psfig]{revtex}
\begin{document}

\begin{titlepage}

*\begin{flushright}
*{\large CU - TP - 836}
*\end{flushright}

\vskip 1\baselineskip
\begin{center}
\baselineskip=24pt
\mbox{}\\[2ex]
{\large \bf Specific Interactions at Cosmic Ray Energies
for Extensive Air Showers Experiments}$^{*}$
\footnotetext{*This work was supported by the Director,
Office of Energy Research, Division of Nuclear Physics of the Office
of High Energy and Nuclear Physics of the U. S. Department of Energy
 under Contract $No. DE-FG02-93ER40764$.} \\ 

{ \bf V.Topor Pop}$^1$ 
\footnotetext{1.  E-MAIL TOPOR@KAON.TAU.AC.IL;
  TOPOR@NT3.PHYS.COLUMBIA.EDU} \\
{\em Physics Department, Columbia University, New York, NY 10027}\\
{\em and}\\
{\em School of Physics and Astronomy, Tel Aviv University,
Tel Aviv, Israel}\\[2ex]
 {\bf M. Gyulassy}\\
{\em Physics Department, Columbia University, New York, NY 10027}\\[2ex]
 { \bf H.Rebel}\\
{\em  Forschungszentrum Karlsruhe, Institut fur Kernphysik III,}\\
 {\em  P. O. Box 3640, D-76021 Karlsruhe, Germany}\\[2ex]

%\today\\
\vskip 0.3cm

{ \it Work submitted to Astroparticle Physics}
\end{center}

\newpage
\vskip 10cm

\begin{abstract}
\normalsize
\baselineskip=24pt

  The HIJING and VENUS models of relativistic hadron-nucleus and
 nucleus-nucleus collisions are used to study interactions of
 hadron-hadron, hadron-nitrogen and nucleus-nitrogen collisions, 
 specific for the extensive air shower
  developments initiated by cosmic rays in the atmosphere.
  The transverse energy, transverse momenta and spectra of
  secondary particles as well as their energy and mass dependence 
  have been  investigated in detail. Results are presented with
 particular emphasis on the contributions of minijets in
 HIJING model  and validity of superposition models 
 in this energy range .\\
 PACS : 13.85.Hd ; 13.85.Ni ; 13.85.Tp ;96.40.De;96.40.Pq;96.40.Tv   
\end{abstract}

\end{titlepage}

\newcommand{\lsim}
{\ \raisebox{2.75pt}{$<$}\hspace{-9.0pt}\raisebox{-2.75pt}{$\sim$}\ }
\newcommand{\gsim}
{\ \raisebox{2.75pt}{$>$}\hspace{-9.0pt}\raisebox{-2.75pt}{$\sim$}\ }
\baselineskip=18pt
\parindent=0.25in
\abovedisplayskip=24pt
\belowdisplayskip=24pt

\newpage

\section{Introduction}

The investigation of the detailed shape of the energy spectrum and the
 mass composition of primary cosmic rays is currently a most active
  field of astrophysical research \cite{rebel0},\cite{rebel1}.

 Experiments on satellites or with balloon-borne detectors give
 information up to energies of ca. $\, 10^{14}$ eV \cite{mul1}. 
 Due to their limitations in size and weight they can hardly be extended 
   beyond$\,\, 10^{15}\,\, $eV, and
   indirect techniques, the observation of the particle-cascades in the
 atmosphere (extensive air showers: EAS), have to be invoked. The
 information about nature and energy of the primary particles is
 reflected by the shower development 
 whose details and signatures for the
 primary particle depend on the high-energy nuclear interactions,
governing the cascading processes. Thus the analysis requires a
 reliable description of these processes, formulated as a hadronic
 interaction model which can be used as generator of Monte-Carlo
  simulations of air showers. It should describe the currently available
 experimental information from accelerator experiments (in particular the
 data from the large collider facilities at CERN and Fermilab) and allow
 a justified extrapolation to experimentally unexplored energy regions.
In the case of the EAS cascades, the quest is for the cross sections
(multiparticle production, rapidity and transverse momentum
 distributions) for hadron-hadron, hadron-nucleus and nucleus-nucleus
collisions as function of the energy (from pion production threshold up
to ultrahigh energies), most importantly for the forward fragmentation
 region, while actually the central region of the collisions is best
 studied at accelerators. The fragmentation region is nevertheless also
 relevant for the interaction models describing the experimental
 observations at SPS energies and beyond 
 \cite{ranf94a,ranf96_2}, \cite{mik95},\cite{topor2}.

 There are many hadronic transport models en vogue which address this
 problem. They comprise versions of the Dual-Parton model (DPM)
 \cite{dpm94}, Quark-Gluon String models (QGSM) \cite{ame1},
 and models designated with the
 name of the code like VENUS  \cite{wer7}, FRITIOF \cite{ander},
 HIJING \cite{wang0}-\cite{wang3} ,Parton Cascade Models 
 \cite{geiger95},\cite{amelin95} and others.
 Some have been specifically developed as Monte-Carlo generator for air
 shower simulations at cosmic ray energies like DPM \cite{capde89},
  HEMAS \cite{forti90} and SYBILL \cite{flet94}.

   Recently an exhaustive comparison of the interaction models with
  each other and with experimental data was performed for primary 
  nucleons, mesons and nuclei in the energy range from 
  $E_{lab} = 10^{11}$ to $10^{17}$ eV colliding with nucleons and 
  nitrogen \cite{knap1,knap3}. 
  VENUS model approach   \cite{schatz1}
 linked to the CORSIKA code \cite{capde92} (now
 widely used for cosmic rays EAS simulations) has been used to
 scrutinize the superposition hypothesis in nucleus-nucleus collisions
 at ultrahigh energies. The hypothesis has been shown to be a rather good
 approximation, for the air shower cascade while Ranft  
 \cite{ranf94b}, \cite{ranf96} using the
 DPMJET-II version of the DPM approach and considering the fragmentation
 and central regions with equal importance concluded that the
superposition is a rather rough approximation of the reality. 
 It was emphasized \cite{knap1} that the comparison cannot
 give preference to one or another of the interaction models and 
 the only way to decide between models is by comparing these with
 experimental data.
 
 Our present work is based on the experience with 
 HIJING and VENUS  models.
 Systematic data \cite{na35_7,marg96} from heavy ion collisions 
 provide new information.
  Both string-models have been applied to a variety of pp, pA and AA
 collision data (see references \cite{topor2}, \cite{wer7},
\cite{wang1},\cite{wang2}, \cite{topor1}).
 However, a consistant
intercomparison of predictions of multiparticle production, transverse
momentum and rapidity distributions at ultrahigh energies and with
respect to their relevance in the EAS cascade is missing. The present
 paper is a first attempt of such a comparison, revealing the most
 salient features and differences in a selected number of cases. We
 introduce the presentation of numerical results with a brief reminder
 of the basis of the HIJING and VENUS  
models under consideration, stressing the
 different procedures in defining the interacting nucleon configurations,
 the quark-gluon string formation and the decay into secondary particles.

 The models have been tested and compared at accelerator energies
 for proton-proton and nucleus-nucleus interactions 
 and then theoretical predictions on pseudorapidity distributions
 of transverse energy, transverse momenta and secondary particles
 spectra as well as their impact parameter, 
energy and mass dependence are given 
 using HIJING model for proton-proton, pion-proton, 
proton - Air Nucleus (p+Air) interactions
 between 1 TeV - 10000 TeV and  for Nucleus - Air (A+Air) interactions
 at 17.86 TeV/Nucleon (corresponding to 1 PeV for iron (Fe) nucleus)
 and between 1 TeV - 10000 TeV. We set for Air nucleus , 
 nitrogen (N) nucleus. 
 A comparison with the results of ref. \cite{knap1}, \cite{schatz1} 
 at the same energy  using VENUS model is also done.
 The  Feynman scaling behaviour of the  model in this energy region,
and effects of the multiple minijets production   
  on  charged multiplicity distributions are investigated.
  Finally a brief discussion on validity of superposition 
  models, taken into consideration mean integrated values 
  of transverse energy and spectrum weighted moments (as defined in 
 references \cite{gaisser90}, \cite{ranf96})
predicted by HIJING model, is presented.

\section{Outline of Theoretical Approach}
 
%\subsection{Outline of  HIJING   Model }

A detailed discussion of the HIJING Monte Carlo model was reported
in references \cite{wang0,wang3}.
The formulation of HIJING was guided by the LUND-FRITIOF and Dual
Parton Model(DPM)  phenomenology for soft nucleus-nucleus reactions at
intermediate energies ($\sqrt{s}<20\,\, GeV$) and implementation 
pQCD processes in the PHYTHIA model \cite{sjos94}
 for hadronic interactions.

Unlike heavy ion collisions
at the existing AGS/BNL and SPS/CERN energies, most of the physical
processes occurring at very early times in the violent collisions of 
heavy nuclei at cosmic ray energies
 involve hard or semihard parton scatterings
which will result in enormous amount of jet production and 
can be described in terms of pQCD.
Assuming  independent production, it has been shown that
the multiple minijets production is important in proton-antiproton
($p\bar{p}$) interactions to account for the increase 
of total cross section \cite{gaisser} and the violation 
of Koba-Nielsen-Olesen (KNO) scaling of the charged 
multiplicity distributions \cite{wang1}.
  
        In high energy heavy ion collisions, minijets 
have been estimated \cite{kaja} to produce 50\% (80\%) of 
the transverse energy in central heavy ion collisions at 
RHIC (LHC) energies. Though they are not identified as single 
distinct jet they can led 
 to a wide variety of correlations, between observables  like
multiplicity, transverse momentum, strangeness and fluctuations 
that compete with the expected signatures of a QGP. This have been
studied in proton-proton ($pp$) or 
antiproton-proton $\bar{p}p$ collisions.

  The HIJING model incorporate nuclear effects such
   as parton shadowing and jet quenching. HIJING is designed
   also to explore the range of possible initial conditions
   that may occur in nuclear collisions at Cosmic Ray
    and colliders  (RHIC,LHC) energies.

 In  high energy heavy ion collisions, a dense
hadronic or partonic matter must be produced in the central 
region. Because this matter can extend over a transverse 
dimension of at least $R_A$, jets with large $P_T$ from 
hard scatterings have to traverse this hot environment. For 
the purpose of studying the property of the dense matter 
created during the nucleus-nucleus collisions, 
HIJING model include an option to model jet quenching in 
terms of a simple gluon splitting mechanism \cite{wang1},\cite{wang2}.

  The main usefulness of these schematic approaches 
 for nuclear shadowing and jet quenching is to test
the sensitivity of the final particle spectra .

The VENUS model is a "colour exchange" (CE) model ,
based on DPM and Gribov-Regge theory (GRT).
A common feature of all Gribov - Regge models is their
formulation within the framework of relativistic quantum theory . The basic
exchange "particles"  of high energy hadron-hadron scattering are
Reggeons and Pomerons. Reggeons and Pomerons are not elementary, 
they are associated with complex diagrams.
A Reggeon is a planar QCD diagram,
a Pomeron a cylindrical QCD diagram, with the planar and cylindrical
surfaces containing networks of gluons and closed quark loops.
A simple model for the Pomeron is a gluon ladder.

HIJING is a minijet model which use 
a critical Pomeron with an intercept $\alpha(0)=1$.
 Within such a minijet scheme, it is assumed that the rise of the 
 cross sections with energy is totally due to 
 to the rise of the minijet cross section .

Elastic amplitudes and hence total cross sections are calculated in VENUS
model from Gribov\'s theory or multiple Pomeron exchanges. Eikonal form is
derived in VENUS \cite{wer7} .
The mechanism for particle production in VENUS is colour exchange with the
weight for ''m'' colour exchanges always being 
$\sigma _m/\sigma _{in}$ where 
$\sigma _m$ are topological cross sections, which 
correspond to cutting ''$m$''
Pomerons and $\sigma _{in}$ is total inelastic cross section.
All GR models  ( VENUS, DPM, QGSM) consider colour exchange (CE) 
 as the basic nucleon-nucleon mechanism whereas 
 $pQCD$ model calculate the cross sections perturbatively 
 using $QCD$ improved parton model.

   In the VENUS model colour exchange in nondiffractive 
   processes applies for valence quarks ( as far as available)
   and sea quarks. A unique feature of the VENUS procedure 
   is the active role of antiquarks which also participate
   in the colour exchange (they are 8 possible diagrams) so that 
   quarks and antiquarks are handled on equal footing. 
   In addition the equivalence of the first CE with all
   further CE's is a unique property. In all other models 
   the first colour exchange is different to the others.
   VENUS is the only model which allows diquark break up,
   with the consequence of formation of {\em "double strings"}.

The basic assumption of the model is that the projectile nucleon -whatever
its status  is after the first collisions - 
moves through the nucleus on a straight forward.
Each interaction implies 
 colour exchange and string formation.
 Such  procedure starting from a
parton level and forming  hadron is refered  as fragmentation or
hadronization.
In the VENUS model the strings are fragmented according the
AMOR procedure (''area law models'') (\cite{wer7})
being based on relativistic classical string dynamics.
 "Area law models" provide a conventent gauge invariant
 energy and momentum string breaking procedure .
Whereas AMOR allows a string to split
into two substring with arbitrary masses 
($string \rightarrow string_{i}+string_{j}$)
the HIJING model which use JETSET algorithm 
requires one of the substrings
to be on shell hadrons ($string \rightarrow string_{i}+hadron$) .

In both approaches the itterative procedure terminates, wherever the string
masses are below a specified cut-off. Those strings are 
identified with stable hadrons or known resonances.

  VENUS allows diquark break up,
   with the consequence of formation of {\em "double strings"},
a non - conventional mechanism which may form when one 
projectile nucleons interacts
with two or more target nucleons. A double string is defined as a colour
singlet baryon configuration consisting of one 
projectile quark connected to
two different valence quarks in the target via a three gluon vertex.
Double strings(one forward quark linked via two normal 
strings to two backward quarks ) do additionally fragment into leading
 baryons and therefore there is not a principal difference 
 from the proton.
However, there is a difference concerning hyperon production.
 The probability in the fragmentation region
is enhanced by a factor of two relative to single string rates(double
strings need two breakups to form a baryon and have therefore 
 a double chance to 
 produce an strange($s$) - antistrange($\overline{s}$) pair ). 
 In the used version(VENUS 4.12) the double string phenomenology is
constrained to reproduce the inclusive   $p+A\rightarrow p+X$ data.

The HIJING model adopts a linear extrapolation  
 of particle production dynamics
 from proton - proton ($pp$) to nucleus-nucleus($AA$)
interactions, taking into account as essential ingredients the 
geometry of the nucleus and kinematical constraints.

Both models (VENUS and HIJING) take into account 
soft and hard scattering.In the
HIJING the hard component is calculated from $pQCD$ 
using a small-transverse  momenta cut off.
 Minijets are accounted in VENUS model by
considering soft and hard Pomerons
(or equivalently soft and hard CE's) ; 
 $\omega =\omega _{soft}+\omega _{hard},$
 where $\omega $ is the Fourier
transform of the Pomeron propagator.
A soft Pomeron provides the standard
colour exchange; the string endpoints acquire 
some intrinsic transverse
momenta ($p_t$) of the order of few hundred MeV.
For hard CE the transverse momentum
$p_t$ of the endpoints is distributed as
$\left( p_t^2+m_h^2\right) ^{-n_h}$
where $m_h$ and $n_h$ are energy independent parameters.

In VENUS model the $pQCD$ results are used and the relative weight of hard
and soft contributions are determined by comparison with data .

\section{NUMERICAL RESULTS}

In the following we present a series of calculated results of the models 
for inter-comparison. For the cosmic ray cascades in the atmosphere,
first of all collisions of protons and heavier ions with
 ${^{14}}N$ and ${^{16}}O$ are important. For these cases there is
 a scarcity of data.
 However,experimental data are of much better quality in hadron-hadron
and especially in proton-proton or antiproton-proton collisions 
 which should be used to   test  the models.

\subsection{ PROTON - PROTON AND NUCLEUS-NUCLEUS 
INTERACTIONS  AT ACCELERATOR ENERGIES}

 We start to present results of $pp$  and $p\,\bar{p}$ collisions .
 We include a comparison of hadron yields
at collider energies ($\sqrt{s}=540\,\, GeV \,\, $)
for $\,\,\bar{p}p\,\,$ interactions, where mini-jet 
production plays a much more important role.
>From different collider experiments
Alner et al. (UA5 Collaboration) \cite{1ua5}  attempted to piece
 together a picture of the composition of a typical soft event at the
 Fermilab $\,Sp\bar{p}S\,$  collider \cite{ward}. 
The measurements were made in various
different kinematic regions and have been extrapolated in the full
transverse momenta($\,p_{T}\,$)  and rapidity range for comparison
as described in reference \cite{1ua5}.
The experimental data are compared to theoretical values obtained with
$HIJING^{(j)}$ in Table I.
 It was stressed  by  Ward \cite{ward} that 
 the data show a substantial excess of photons compared to the
 mean  value for pions $\,\,<\pi^{+}+\pi^{-}>\,\,$. 
 It was suggested as a possible 
 explanation of such enhancement a  gluon 
 Cerenkov radiation emission in hadronic collision
\cite{drem}. Our calculations rules out such hypothesis.
 Taking into account decay from resonances and 
 direct gamma production,  good 
agreement is found within the experimental errors.

In the following plots
the kinematic  variable used to  describe 
single particle properties are
 the transverse momentum $\,p_{T}\,$ and the rapidity $\,y\,$ defined 
as usual as:
 \vskip 0.1cm
 \begin{equation}
 y=\frac{1}{2}ln \frac{E+p_{3}}{E-p_{3}}=ln\frac{E+p_{3}}{m_{T}}
 \label{e18}
 \end{equation}
 \vskip 0.1cm
 with $\,E,\,\,p_{3}\,$, and $\,m_{T}\,$ being energy, 
longitudinal momentum and transverse mass
$ m_{T}=\sqrt{m_{0}^{2}+p_{T}^{2}}$ with $\,m_{0}\,$ 
being the particle rest mass.

  The pseudorapidity $ \eta $ is used rather than the rapidity 
  since for $ \eta $  no knowledge of particle masses is 
  required :
 \vskip 0.1cm
 \begin{equation}
 \eta = \frac{1}{2}ln \frac{p+p_{3}}{p-p_{3}}= - 
 ln tan{\frac{\theta}{2}}
 \label{e19}
 \end{equation}
 \vskip 0.3cm
  where $p$ is  the projectile nucleon momentum and $\theta$ is the
  scattering angle.

 Feynman $x_F$ variable is defined for ultrahigh energy as:
 \vskip 0.1cm
 \begin{equation}
 x_F=2 \frac{m_T}{\sqrt{s}}sinh(y^{cm})
 \label{e20}
 \end{equation}
 \vskip 0.1cm
  where  $y^{cm}$  ans $s$ are  rapidity 
  and total energy in center of mass frame (cms).

 In Figure 1 we compare rapidity and transverse momentum
       distributions for strange particles in proton-proton
      interactions at 300 GeV  given by HIJING model with
   experimental data \cite{ex80}. The agreement is quite good. 
   However, it will  be interesting to investigate also
   the Feynman scaling behaviour of the model at the 
   accelerator energies since the forward fragmentation
   region seems to play an important role \cite{ranf96},
   \cite{topor2}.

    We foccussed our analysis concerning particle production 
    in nucleus - nucleus interactions  at SPS energies,
    where the models should be better tested, mainly for 
     interactions $S+A$ at 200A GeV and  $Pb+Pb$
    at 160A GeV. 
    
    Figure 2 shows that the hyperon production data require new
    mechanisms not apparent in proton-proton reactions (see Fig. 2a,
    Fig. 2b). Recall that VENUS has an extra degree of freedom
   relative to HIJING, a color rope effect, called 
  {\em double string}. The factor of two enhancement of hyperon
  production is already needed (see dotted histogram). Adding 
  final state interactions VENUS should describe better hyperon
 yields in AA and pA ( see ref. \cite{topor2}) interaction
 ( see Fig. 2a,b,d - dashed histograms). 
  The excess rapidity shift of the hyperons 
  relative to non-strange baryons is also clear from Fig. 2a and 
  Fig. 2c ,where in Fig. 2a the peaks are shifted approximately one 
 unit of rapidity further than those calculated for protons
  \cite{topor2}. The enhanced hyperon transport is currently 
  not explained by any of these models.

  The rapidity distributions for antiproton  $\,\bar{p}\,$
  and for negative kaons $K^{-}$ are represented for 
  $S+S$ interactions at 200A GeV in Figure 3 
  (Fig. 3a ($\,\bar{p}\,$) and Fig. 3b ($K^{-}$)). We see that 
the effect of including extra degree of freedom and 
 final state interaction is less pronounciated . 
 Therefore we give theoretical predictions  for  ($\,\bar{p}\,$)
(Fig.3c) and ($K^{-}$)) rapidity distributions (Fig.3d) for $Pb+Pb$ 
interactions  at 160A GeV without final state interactions in VENUS model
 comparing with HIJING.
 
We turn next to the distribution of valence baryons
in $A+A$ collisions at SPS for which data have
finally become available from NA49\cite{na49_4,na49_1} and
NA44\cite{xu_96} and provide the most important test of the models.
 As noted in \cite{mik95} data on
nuclear stopping power in $Pb+Pb$ collisions  have been long awaited
as a critical test of nuclear transport models and with ongoing search
for nonlinear dynamical phenomena in nuclear reactions. One source 
of nonlinear behavior may arise if a quark - gluon plasma is 
formed in such reactions. 

Figure 4 compares the spectrum of pion and participants protons
in $S+S$ at 200 AGeV  \cite{na35_4} and $Pb+Pb$ reactions
at $160$ AGeV \cite{na49_4,xu_96,na49_1}.
The various data sets from NA35 for $S+S\rightarrow \pi^- + X$
correspond to different centrality triggers \cite{na35_4},
with the higher one corresponding to a more severe veto trigger cut
 (see reference \cite{na35_4} for more details).
We see that the negative pion rapidity densities
are well accounted for by both
 HIJING  and VENUS models.
This is largely due to the fact that the pion distribution
simply grows linearly with the atomic number between $S+S$ and $Pb+Pb$.
The centrality trigger was implemented in the above calculations via an
impact parameter cut of $b<1$ fm.

The large central hole in the $dN/dy$ predicted by HIJING
is traced back here to the central hole in $pp$
in FRITIOF type models. By choosing fragmentation functions
that fit the $pp$ central rapidity
region better, VENUS, can avoid the strong suppression
of the mid-rapidity protons of HIJING and other  models
using JETSET. We note that HIJING 1.3 utilizes the new particle
conventions of JETSET 7.2 . In this version it is no longer 
possible to vary the diquark fragmentation scheme.
The analysis that we have performed points to the necessity of 
modifying the baryon fragmentation part of HIJING, FRITIOF 
and DPM models.  

Figures 4a,c demonstrate also the insensitivity
of the pion distribution to the underlying baryon number flow.
In particular, the rather large difference between
the proton distribution in HIJING and VENUS in Fig. 4d is
contrasted by the much more modest difference of the pion
distributions in Fig. 4c. 
 It was  shown \cite{topor2} that
the forward energy flux is a more sensitive measure of the
energy degradation difference between the models.

\subsection{PROTON -AIR NUCLEUS INTERACTIONS 
 AT COSMIC RAYS ENERGIES}

 There are not sufficient experimental data available for the
 fragmentation region of hadron collisions with light target 
 nuclei. Hence many feature for these collisions have to be
 infered by studies of interaction models. 

 A comparative study of the predictions of different models
 has been done recently  in ref. \cite{knap1,knap3}. 
It was shown that clear differences 
 in the results exist and that differences between models are
 smaller for iron primaries than for primary protons.
 Since consistant theoretical predictions of 
multiparticle production with respect to their relevance to EAS 
cascade is missing for HIJING , we performe such kind of analysis. 
 Some specific interactions were investigated
 in HIJING approach at RICH and LHC energies \cite{wang1}-\cite{wang3}.

   We start  taken  into consideration some specific spectra of 
   secondary particles analysing in detail especially 
   transverse energy, transverse momenta or Feynman $x_{F}$ 
    distributions.

  We generate for proton-Air Nucleus ($p+Air$)
  interactions $10^4$ minimum bias events 
 ($b_{min}=0, \,\,b_{max}=5 fm$).
  In order to give an idea about energy used for producing 
  new particle we investigate in Figure 5 the transverse 
  energy pseudorapidities spectra and their 
  dependence with energy for all secondaries (Fig. 5a),
  all neutral (Figure 5b) ,all charged (Fig. 5c) and
  gluons (Fig. 5d). As we see from Figure 5  gluons 
  carried an important fraction from transverse energy 
  and this fraction increase with increasing energy 
  (from  2.8 \% at 1 TeV to  17.3 \% at 1000 TeV).
   Also the percentage of occurence of gluons increase
   from 7.20 \% at 17.86 TeV to 10.71 \% at 1000 TeV 
   (see Table II).

     The rapidity distributions and energy dependence 
     of main secondary particles are shown
     in Figure 6 and Figure 7. Secondary $\pi^{\pm}$ (Fig. 6a and 6b)
     and $K^{\pm}$ (Fig. 6c and 6d) are the source of Cosmic Ray Muons
     and the source of atmospheric Neutrinos produced by the 
     Cosmic Ray cascade. Secondary $\pi^{0}$ (Fig. 7a) and $\eta$ mesons
     are the source of the electromagnetic shower and secondary 
     charmed mesons (Fig. 7d) are the source for prompt Muons 
     and Neutrinos . The lambdas have been shown (Fig. 7c), because 
     they can be produced from protons by exchanging a single valence 
     quark by strange quark. The outer maximum of their distributions
     are due to this process. This component is identified in all
     nucleon distributions (neutron, proton). 
     The statistics ($ 10^{4}$ events generated) seems not to be 
     enough for charmed mesons ($D^{\pm}$) but we give such 
     distributions only to show that introducing HIJING code in
     a shower code with a much higher statistics at simulation
     level a study of promt muon component should be feasible.

     We see from Figure 6 that mesons distributions exhibit a 
     broad structureless shape which does not depend strongly 
     on the type of mesons but have a dependence on energy 
     from   1 TeV to 1000 TeV.

     Trying to stress out  the relevance of accelerator data 
      on particle production in hadron - nucleus collisions for 
      Cosmic Ray cascade we study the Feynman scaling behaviour of 
      $p+Air \rightarrow \pi^{\pm} + X $  and 
      $p+Air \rightarrow p+\bar{p} + X $ in Fig. 9a and Fig. 9b 
      respectively.
      We plot the $x_{F}dN/dx_{F}$ distributions for laboratory
      energies of 1 TeV (dotted histograms), 100 TeV (dashed 
      histograms), 1000 TeV (solid histograms).
      The violations of Feynman scaling which occur 
      are connected with 
      known rise of rapidity plateau for all kinds of produced 
      particles and with production of minijets.
       Due to minijets Feynman scaling is more strongly 
       violated especially in the region $x_{F} \geq 0$ .
       The violation of Feynman scaling are less dramatic in
       DPMJET II model \cite{ranf94a},\cite{ranf94b} and appear only
       around $x_{F}=0$ and $x_{F}=1$. We note also that HIJING 
      show violation of KNO scaling due to the production of 
      multiple minijets and the tendency becomes stronger with 
       increasing energy \cite{wang1}.

      In order to evaluate multiplicity distributions for  the 
      charged particles in $p+Air$ interactions (Fig. 10a)
      we differentiate  the contributions from soft 
      ( events with $N_{jet}=0$  represented in Fig. 10b) and 
   hard processes ( events with $N_{jet}=1$  in Fig. 10c and for  
       events with $N_{jet}> 1$  in Fig. 10d) where $N_{jet}$ is
      the number of minijets produced in that events.
       Our calculations including the effects of multiple 
       minijets  are the contributions from the events which 
       have hard collisions with $P_{T} > P_{0j}$ = 2 GeV/c. 
  Analysing  Figure 10 it is clear that the events at the tails of 
       the charged multiplicity distributions 
       in p+Air interactions are mainly those with
       multiple minijets production.

 \subsection{NUCLEUS-NUCLEUS AIR  INTERACTIONS AT COSMIC RAYS ENERGIES}

  Nucleus-Air collisions are of great importance in the
  EAS development. It is important, that the model will be
  able to give a good description of hadron production in
  nucleus-nucleus interactions. In this subsection we
  try to investigate mainly the dependence on projectile mass
  for specific interactions for EAS (He, Ne, S, Fe+Air Nucleus)
  at 17.86 TeV/Nucleon which correspond to 1 PeV laboratory
  energy for Fe nucleus.

   At SPS energies the calculated number
  of target and projectile participants as well as the number
  of participants as a function of reaction impact parameter
   (b fm)\cite{awes89} shows  no difference 
   in HIJING and VENUS model \cite{topor2}.

 The results depicted in Figure 11 are obtained 
 for $10^4$ generated events and for the following
 intervals  of impact parameter ($b_{min}-b_{max}$) : 
  Fe + Air (0-13 fm), S + Air (0-11 fm), Ne + Air (0-10 fm),
  He + Air (0-7 fm) - dotted histograms ;
  Fe + Air (0- 8 fm), S + Air (0-7 fm), Ne + Air (0-6 fm), 
He + Air (0-5 fm ) - dashed histograms ;
  Fe + Air (0 - 5 fm), S + Air (0 - 5 fm), Ne + Air (0 - 5 fm), 
He + Air (0-4 fm) - solid histograms.
   
    Figure 11(a, b, c, d) shows strongly dependence on 
 impact parameter intervals ($b_{min}$-$b_{max}$) for theoretical 
 predictions at 17.86 TeV/Nucleon .

  Therefore for comparison of HIJING and VENUS results at 
   17.86 TeV/Nucleon for p+Air, He+Air,
  Ne+Air, S+Air, Fe+Air and  for p+Air at 
  1000 TeV we try to get approximatively the same number 
  of participants. For HIJING model we get the values 
  for mean numbers of participants listed in last lines
  of Table II for $10^4$ generated events and 
   the following impact parameter intervals ($b_{min}-b_{max}$):
  He+Air(0-4 fm), Ne+Air(0-6 fm), S+Air (0-7 fm), Fe+Air (0-8 fm) ,
   $p+Air$ (0- 5 fm).  The results for VENUS model are taken from 
  Schatz et al. \cite{schatz1}.
  
  Table II lists the frequency of various particles among the 
  secondaries for collisions of different nuclei with 
  nitrogen. Our calculations confirm the results from
  reference \cite{schatz1} .The percentages do not seem to 
  depend strongly on projectile mass nor, as shown by proton 
  results on energy. We see only a slight tendency of 
  increasing strangeness production with increasing primary
  mass for VENUS results.
  Analysing the results of Table II  we see that the VENUS model
  predict  more pions and kaons, but less gamma particles that 
  HIJING model . We remark also differences in total multiplicities
  at ultrahigh energies which can not be explained only by 
   the difference between total number of participants .
   It seems that  "double string" mechanism change considerably
   the baryon spectra and allows the baryon number to migrate 
   several units of rapidity from the end point rapidity.

   Important for Cosmic Ray studies and EAS development is 
   also the dependence of particle production on the nuclear 
   target and projectile. Taking into consideration the same
   conditions as those reported for the values listed in 
   Table II the theoretical predictions for
  mass dependence of pseudorapidities spectra  
   are given  for transverse energy in Figure 12,
  for transverse momenta in Figure 13 and
  for main secondaries produced new particles in Figure 14.
  The  transverse momenta  distributions of all charged,
  proton, $\pi^+$,$\pi^-$ are displayed in Figure 15 for 
  limited interval of pseudorapidity $ |\eta| < 0.5$.

  For Feynman $x_{F}dN/dx_{F} $ distributions 
   we give theoretical values for specific EAS interactions 
  in Figure 16 only for
   charged pions( Fig.16a) and for  protons(Fig.16b). 
    We see a slight mass dependence of Feynman distributions
    and transverse momenta  distributions for $A \geq 20$
    (Fig.15a-d).
 
    Instead of the proper sampling of Nucleus + Air Nucleus
   scattering events, an approximation often applied in
   EAS development is the so called superposition model.
   There are two different possible superposition models:
   a nucleus-nucleus collision A-B with $N_{part}$
   participating nucleons is approximated as the superposition
   of $N_{part}$ simultaneous nucleon-nucleon collisions and
    the second one : a nucleus-nucleus collision  A-B  with
    $N_{proj}$ participating projectile nucleons is
     approximated as the superposition of $N_{proj}$ simultaneous
     nucleon-B collisions. The validity of this principle
     was analysed  in some recent works \cite{schatz1},
     \cite{ranf94a},\cite{ranf94b},\cite{ranf96} 
     and older one \cite{serge89},\cite{serge93} 
    with different conclusions.
     
    In this section we investigate using HIJING approach the integrated 
    mean transverse energy for secondaries
     produced in Nucleus+Air Nucleus interactions
     at 17.86 TeV/Nucleon. So, we  generate $10^4$ events in the same
     impact parameter interval (0-5 fm) for Fe, S, Ne, He, p+Air
      interactions. The values obtained for mean projectile
     participants ($N_{proj}$) and mean number of
      binary collisions  $N_{coll}$(which include nucleon-nucleon($N-N$),
      nucleon-wounded nucleon($N-N_w$),
     wounded nucleon-nucleon $N_w-N$ and 
    wounded nucleon-wounded nucleon $N_w-N_w$ 
      collisions) are listed in Table III. 
     The results given in Table III  are for all secondaries ,
     all charged and all neutrals particles produced in
     interactions. The  values of integrated
     mean transverse energy $<E_T>$
     scale with the number of binary collisions in nucleus-nucleus
     interactions at this energy  $N_{coll}$ and scaling propreties
      are valid for $A \geq 20$ and should not be
      applied to light nucleus+Air interactions.
      Since  integrated mean value of transverse energy $<E_T>$
       is a measurable quantity it will
      be interesting to verify this scaling at RHIC
    and LHC  energies.

  In order to evaluate multiplicitity distributions for 
  charged particles in A + Air interactions we differentiate
  the contribution for all events (soft + hard)
  (Fig. 17a for Fe + Air and Fig. 17b for S + Air ) 
   and only hard processes - events with 
  $N_{jet} > 1 $ (Fig. 17c for Fe + Air and Fig. 17d for
  S + Air ). We can see from Figure 17 that the low multiplicity 
  events are dominated by those of no jet production while
  high multiplicity events are dominated by those of at least 
  one jet production.
   Also it is clear from Figure 17 that the contributions
  from the events which have hard collisions  increase
  with increasing of available energy and of projectile mass.

   At ultrahigh energies   nuclear effects like 
  nuclear shadowing of partons and jet quenching (see section 2),
   should have important contributions \cite{wang1},\cite{wang2}.
   In HIJING model a simple parametrization of gluon shadowing 
   and a schematic quenching model 
   were introduced to test the sensitivity 
  of the final distributions to these aspects of nuclear 
  dynamics.
  Trying to estimate the results with and without  such effects 
  we investigate the
  differences for more central Fe + Air interactions 
  at fixed energy 17.86 TeV/Nucleon.
   Our calculations seems to indicate  for main secondary 
   particles produced in EAS specific interactions 
     in this energy 
   region, that the effects should be neglected (they are less than 
   2 -3 \%).

  \section{More Specific Spectra at Cosmic Ray Energies}

   \subsection{Feynman scaling}
  
   Since experimental data in the projectile 
   fragmentation region are of much better quality in hadron-hadron
   collisions,
   we study the Feynman scaling behaviour of the model 
  for energy range well below 1 TeV. 
  In Fig. 18a a one compare results for $\pi^{-}$ production in
  the forward and backward fragmentation regions
  for proton - proton interactions at 250 GeV
  \cite{adam88} and find a reasonable agreement.
  In Fig. 18b  the theoretical
  predictions are compared with experimental data from
  Bailly et al. \cite{bail87} at 360 GeV (Fig. 18b) for $\pi^{-}$
  and from
  Aguilar-Benitez et al. \cite{agui91} at 400 GeV for $K^{+}$ (Fig. 20c).
  In Fig. 18d the data for $\pi^{+}$ production at 205 GeV 
 \cite{kaf77} are also compared with HIJING model prediction.
  All sets of data strongly disagree with the model predictions.
  Furthermore , the data of Aguilar-Benitez et al. \cite{agui91}
  shows a strange structure at $0.5 \leq x_{F} \leq 0.8$ 
  which is not found in other experiments and in the model
  \cite{ranf94a}, \cite{ranf94b}.
  We note that the model is well below the data for 
  $p + p \rightarrow K^{+} + X$ and for 
  $p + p \rightarrow \pi^{+} + X$.

  In Figs. 19a,b we compare $x_{F}d\sigma/dx_{F}$ distributions 
 for $\Lambda$ hyperons and neutral mesons $K^{0}_{s}$ with
 the experimental data of Bailly et al., at 360 GeV 
 \cite{bail87,bail87} . The agreement of model and data are quite good for
 $K^{0}_{s}$ , but theoretical predictions overestimate the data for
 $\Lambda$ in the projectile fragmentation region . 
 Also shown dependence on energy for 
$p + p \rightarrow \Lambda^{0} + X$ (Fig. 19c) and 
$ p + p \rightarrow \gamma + X $  (Fig. 19d) between 1 TeV - 1000 TeV.

  In the cosmic ray cascade the interactions of secondary hadrons
 mainly pions and kaons are as important as interactions of nucleons.

 In Figs. 20a-d we compare the model predictions with the results 
 of the EHS-NA22 experiment \cite{adam88} in $ \pi^{+} + p $
 collisions at 250 GeV for $pi^{-}$ (Fig. 20a) and 
 $h^{+}$ (Fig. 20c ) and with the data of Brenner et al., at 175 GeV 
 \cite{bren82} for $ \pi^{+}$ (Fig. 20b) and for $pi^{-}$
 (Fig. 20d) . 
   These comparisons shows only partial agreement with the data,
  especially in central region.

 \subsection{ Comparison of nucleus - nucleus collisions 
 according to HIJING model with superposition model}
 In section IIIC we have investigated the superposition model
 for transverse energy. In this section we will extend the 
 energy region and also we will investigate inelasticities and
 spectrum-weighted  moments calculated using HIJING model 
 for C-Air collisions. 
 The spectrum-weighted  moments are defined as in Ranft et al.
 \cite{ranf96}.
 We introduce new variables $x_{pL}=p_L^i/P_0$ or 
 $x_{E}=E_i/E_0$ in the laboratory  frame following the basic 
 discussion of ref. \cite{gaisser90}.
$E_i$ ($p_L^i$) is the laboratory energy(momenta) 
of secondary particle i and $E_0$ ($P_0$) is the 
laboratory energy (momenta) of the projectile in hadron-nucleus 
 collisions , or the energy per nucleon in nucleus-nucleus 
 collisions. These variables are similarly to 
 Feynman $x_{F}$, but expressed in the laboratory
 frame. 
 We can consider  $x_{E}$ distributions 
 $F(x_{E})= x_{E} \frac{dN_i}{dx_{E}}$. The cosmic ray
spectrum weighted moments in nucleus-nucleus collisions 
are defined as moments of the $F(x_{E})$ , taking into 
account the power of the integral cosmic ray energy 
spectrum ($\gamma \simeq 1.7$) :

\begin{equation}
 Z_i=\int_{0}^{A_A}\,\,(x_E)^{\gamma -1}\,\, F_i(x_{E})\,\,dx_{E}
\end{equation}

where $A_A$ is the mass number of the projectile nucleus A.

 Table IV and V compare 
 multiplicities and spectrum - weighted moments calculated 
 using HIJING model for C - Air collisions.
  The comparisons in these two tables with 
  both version of 
 the superposition model show, that the superposition is 
 a very rough and unreliable approximation to real nucleus - 
 nucleus collisions.

 In Table VI the inelasticities $K_{h}$ in proton - Air collisions
 calculated from HIJING model are given for the most important 
 secondaries as function of energy.
 Inelasticities are defined in our calculations as the fraction 
 of the laboratory energy carried away in the average by secondary 
 hadron of kind h . From Table VI we see that all values for 
  inelasticities decrease with increasing energies.
   
 In Table VII we give the spectrum weighted moments 
 for pion and kaon production.
 $ Z_{\pi}$ and $Z_{K}$ obtained in HIJING model 
 for  $pN$ collisions in comparison with  the values predicted by   
 others model  : DPMJET \cite{ranf94a} ,\cite{ranf94b}
 HEMAS \cite {forti90} and SIBYLL \cite{flet94}.
  We see that the values in HIJING models have a week dependence 
  on energy from 1 TeV up to 10000 TeV and the moments for 
 $p + N$ collisions are smaller than in $p + p$ collisions .
 We mention that the same result have been obtained in DPM approach 
 by Ranft \cite{ranf94a},\cite{ranf94b}.

\section{Conclusions}

In this paper an  analysis  of  particle 
production is performed for hadron-hadron, 
hadron-nitrogen and 
nucleus-nitrogen collisions, specific for EAS developments
initiated by cosmic rays in the atmosphere.
The models HIJING and VENUS have been used for guidance in 
understanding of primary interactions. They are compared
 with the results at accelerator energies,
in cases where experimental data exist.  

 A very good agreement is found within experimental 
 errors for ultrahigh energies ($\sqrt{s}=540\,\,GeV$)  
 in HIJING approach where mini-jet production plays a 
 much more important role. For nucleus-nucleus collisions 
 at SPS energies the rapidity spectra are well 
  accounted for both HIJING and VENUS models for mesons, but
  VENUS model seems to give a better description for 
   hyperons  production. By choosing fragmentation functions
 that fit the $pp$ central rapidity region better, VENUS, can
 avoid the strong suppression of the mid-rapidity protons of 
  HIJING model. The analysis that we have performed points to
  the necessity of modifying the baryon fragmentation part 
  of HIJING model. At SPS energies a final state 
   interactions in dense matter as well as a color rope 
   effect predicts much higher degree of baryon stopping 
   at midrapidity than HIJING.

  The event generator VENUS \cite{knap1}, \cite{schatz1} 
   and HIJING in the 
  present study were tested to simulate ultrahigh energy
  collisions for hadron-Nitrogen and Nucleus-Nitrogen interactions.  
  The transverse energy, transverse momenta and secondary 
 particles produced spectra, inelasticities, spectrum 
 weighted moments $Z_{h}$, as well as their energy 
  and mass dependences  have been investigated in detail.
 The theoretical predictions for HIJING model suggest a scaling with  
  number of  binary nucleon-nucleon collisions
  for mean transverse energy of secondaries . It will be 
  interesting to verify this hypothesis at RHIC energies and 
   LHC energies. 

 The contributions from the events which have hard 
 collisions are strongly 
 evidentiated at the tail of charged particles multiplicity 
 distributions and increase with increasing energy and 
 increasing mass of projectile.

 At accelerator energies HIJING model gives only partial 
 description of experimental data for Feynman distributions
  and fails to reproduce the fragmentation regions. 
 Feynman scaling is more strongly violated in HIJING model
 due to multiple minijets events, especially in the projectile 
 fragmentation region.
 For Cosmic Ray energies  and specific EAS interactions 
 effects of  parton shadowing and jet quenching  could 
 be neglected. 
 
 Possible utilization inside a shower code of these models
 allows to extend the analysis and to release the simple
 superposition model which seems to be partially not satisfied
 in this energy interval.

\section{Acknowledgements}

We are grateful to K. Werner for providing 
us the source code of VENUS.
One of the authors (VTP) would like to  express  his
 gratitude to M. Morando and R. A. Ricci  for kind hospitality and
 acknowledge financial support from INFN-Sezione di Padova,
 Italy where part of the presented calculations have been performed.

%\newpage

%\newpage

\vskip 0.5cm
\begin{table}
\caption{Particle composition of $p+\bar{p}$ interactions
    at 540 GeV in cms.}
\label{tab1}
\vskip 0.5cm
\begin{tabular}{||c||c|c|c||}  
\hline\hline
{\bf Particle type} & ${\bf <n>}$  & $ {\bf Exp.data}$
& $ {\bf HIJING^{(j)}}$  \\
\hline
\hline
{\bf All charged} &  $29.4 \pm 0.3$ &\cite{1ua5} &  $28.2$  \\
\hline
 ${\bf K^{0}+\bar{K^{0}}}$ & $2.24 \pm 0.16$ & \cite{1ua5} & $1.98$ \\
\hline
 ${\bf K^{+}+K^{-}}$ &$ 2.24 \pm 0.16 $ & \cite{1ua5} & $2.06$ \\
\hline
 ${\bf p+\bar{p}}$ & $1.45 \pm 0.15$  & \cite{ward} &  $1.55$ \\
\hline
 ${\bf \Lambda+\bar{\Lambda}}$ & $0.53 \pm 0.11$  
 & \cite{1ua5} & $0.50$ \\
\hline
  ${\bf \Sigma^{+}+\Sigma^{-}+\bar{\Sigma^{+}}+\bar{\Sigma^{-}}}$  &
  $0.27 \pm 0.06$  &    \cite{ward}    & $0.23$  \\
\hline
   ${\bf \Xi^{-}}$  &  $0.04 \pm 0.01$ & \cite{1ua5} & $0.037$  \\
\hline
  ${\bf \gamma}$  & $33 \pm 3$  & \cite{1ua5} &  $29.02$  \\
\hline
   ${\bf \pi^{+}+\pi^{-}}$  &  $23.9 \pm 0.4$ &\cite{1ua5} &   $23.29$ \\
\hline
   ${\bf K_{s}^{0}}$ &  $1.1 \pm 0.1$& \cite{1ua5}  &   $0.99$  \\
\hline
   ${\bf \pi^{0}}$ & $11.0\pm 0.4$  & \cite{ward} &    $13.36$ \\
\hline
\hline
\end{tabular}
\end{table}

\newpage
\begin{table}
\label{Tab2}
\caption{Percentage of occurence of various particles among 
 the secondaries of a Nucleus-Air collision as calculeted 
 by the HIJING and VENUS models.The number of
 protons and neutrons have been reduced by respective 
 numbers in the primary system(values labeled by star(*)
 \protect{\cite{schatz1}}). The average multiplicity and
the  numbers of participants are also given.}
%\vskip 0.3cm
\begin{tabular}{||c||c|c|c|c|c|c|c||}  \hline \hline
  ${\bf Particle}$ &${\bf Projectile}$ &${ \bf  {^5}{^6}Fe}$ &
  $ { \bf {^3}{^2}S}$ & ${ \bf {^2}{^0}Ne}$
  &${ \bf {^4}He }$ & ${\bf p}$ & ${\bf p}$\\
   ${\bf type} $ & ${\bf Energy(TeV/N)}$ &$ 17.86$ & $ 17.86$
  &$ 17.86$&$ 17.86 $ & $17.86$ & $1000$ \\
\hline
  ${\bf <\pi^-+\pi^+>}$ &$ {\bf HIJING}$ & $ 45.98$ &$45.93$ &
      $ 45.96$ & $45.86$  & $ 45.76$ & $46.57$ \\
      &${\bf VENUS}$ &$ 51.02$ &$ 51.29$ &$51.48$ &$52.34 $
       & $53.05$ & $52.15$ \\
\hline
 ${\bf <\pi^0>}$ & $ {\bf HIJING}$ & $26.13$ & $26.02$ & $26.10$ &
   $25.93$ & $26.04$ & $26.38$ \\
   &${\bf VENUS}$ &$28.30$ & $28.52$ & $28.49$ & $28.85$ & $28.93$ &
   $28.43$ \\
\hline
   ${\bf <K^++K^->}$ & $ {\bf HIJING}$ & $5.20$ & $5.20$ & $5.20$ &
   $5.20$ & $5.16$ & $5.58$ \\
   $ {\bf K \,\,mesons}  $ &${\bf VENUS}$ &$12.35$ & 12.02 & 11.84 & 
    11.11 &  10.51 & 10.86 \\
\hline
    ${\bf <K^0_s>}$ & $ {\bf HIJING}$ & $2.54$ &$2.57$ & $2.53$ &
    $2.52$ & $2.57$ & $2.75$  \\
\hline
${\bf <p> }$ & $ {\bf HIJING}$ & $4.16$ & $4.20$ & $4.20$ &
   $4.37$ & $4.78$ & $3.25$  \\
   ${\bf <p> *}$ & ${\bf VENUS}$ & $0.70$ & $0.66$ & $0.66$ &
   $0.66$ & $-0.11$ & $0.66$ \\
\hline
${\bf <n> }$ & $ {\bf HIJING}$ & $4.12$ & $4.20$ & $4.23$ &
  $4.40$ & $4.16$ & $2.86$  \\
  ${\bf <n> *}$ & ${\bf VENUS}$ & $0.54$ & $0.61$ & $0.64$ &
  $0.66$ & $1.49$ &$ 1.43$  \\
\hline
${\bf <\Lambda>}$ & $ {\bf HIJING}$ & $0.70$ & $0.70$ &$0.71$ & 
 $0.72$ & $0.72$ & $0.56$ \\
 ${\bf <\Lambda+\Sigma^0>}$ & ${\bf VENUS}$ 
 & $1.66$ & $1.64$ &$1.67$ &
  $1.60$ & $1.55$ & $1.31$  \\
\hline
${\bf <\gamma>}$ & ${\bf HIJING}$ & $4.30$ & $4.30$ & $4.33$ &
  $4.23$ & $4.27$ & $4.48$    \\
        &${\bf VENUS}$ & $1.60$ & $1.48$ & $1.43$ & $1.20$ &
        $1.12$  & $1.18$  \\
\hline
 ${\bf all\,\, charged}$ & $ {\bf HIJING}$ & $57.30$ 
 & $57.47$ & $57.54$ &$57.54$ & $57.84$ & $57.79$ \\
\hline
 ${\bf all\,\, neutrals}$ & ${\bf HIJING}$ & $42.33$
 & $42.57$ & $42.61$ &
    $42.50$ & $42.40$ & $42.18$  \\
\hline
 ${\bf <\bar{p}>}$ & ${\bf HIJING}$ & $1.42$ & $1.38$ & $1.43$ & 
 $1.40$ & $1.41$ & $1.56$  \\
\hline
${\bf <\bar{n}>}$ & ${\bf HIJING}$ & $1.40$ & $1.38$ & $1.43$ &
 $1.40$ & $1.41$ & $1.60$ \\
\hline
 ${\bf <q+\bar{q}>}$ & ${\bf HIJING}$ & $0.40$ & $0.43$ & $0.41$ &
 $0.36$ & $0.33$ & $0.71$ \\
\hline
 ${\bf Mean}$ &${\bf HIJING}$ & $274.0$ & $217..5$ & $203.0$ &
 $88.7$ & $38.9$ & $51.7$  \\
 ${\bf multiplicity}$ & ${\bf VENUS}$ & $354.7$ & $270.5$ & 
 $209.3$ & $92.6$ & $49.4$ & $106.5$ \\
\hline
 ${\bf Mean\,\, projectile}$ &${\bf HIJING}$ & $9.6$ & $7.2$ & $6.5$ &
 $2.2$ & $1.0$ & $1.0$ \\
 ${\bf participants}$ & ${\bf VENUS}$ & $12.1$ & $8.1$ & $5.7$ &
 $2.1$ & $1.0$ & $1.0$ \\
 \hline
 ${\bf Mean\,\, target}$ & ${\bf HIJING}$ & $5.9$ & $5.6$ & $5.6$ &
   $3.4$ & $2.06$ & $1.42$ \\
 ${\bf participants}$ & ${\bf VENUS}$ & $6.0$ & $5.3$ & $4.7$ & 
  $3.2$ & $2.0$ & $2.06$ \\
\hline
\end{tabular}
\end{table}

\begin{table}
\caption{Mean transverse energy for  
 the secondaries of a Nucleus-Air collisions 
 at 17.86 TeV/Nucleon as calculeted 
 by the HIJING  model and by considering superposition 
 of $N_{coll}$ nucleon-nucleus collisions and $N_{proj}$
 nucleon-nucleus collisions,where $N_{coll}$ is the number 
 of binary collisions and $N_{proj}$ is the number of 
 participant projectile nucleons. $E_{T}^{pA}$ is 
 mean transverse energy in p+Air interaction at the same 
 energy and in the same impact parameter interval
 (see the text for explanation).}
\label{Tab3}
\vskip 0.5cm
\begin{tabular}{||c||c|c|c|c|c|c||}  \hline \hline
        &${\bf Projectile}$ &${ \bf  {^5}{^6}Fe}$ &
  $ { \bf {^3}{^2}S}$ & ${ \bf {^2}{^0}Ne}$
  &${ \bf {^4}He }$ & ${\bf p}$  \\
\hline
\hline
      ${\bf Mean \,\,number}$ & $<N_{coll}>$ & $27.75$ & $19.21$ & 
      $13.61$ & $2.62$ &   \\
\hline
\hline
 ${\bf Mean\,\, number}$   & $<N_{proj}>$ & $18.43$ & $11.44$ & 
        $8.24$ & $1.73$ &  \\
\hline
\hline
      ${\bf  Mean\,\, Transverse}$ &$ {\bf all\,\, secondaries}$ & 
   $212.5$ & $ 140.7$ & $104.2$  & $ 25.16$ & $7.54$ \\
       $\bf Energy$ &${\bf N_{coll}*E_{T}^{pA}}$ &
     $ 209.2$ &$ 144.8$ &$102.6$ &$19.75 $ &    \\
       $ {\bf (GeV)}$   &${\bf N_{proj}*E_{T}^{pA}}$ &
     $138.96$ & $86.25$ & $ 62.10$ & $13.04$ &  \\       
\hline
\hline
 ${\bf  Mean\,\, Transverse}$ &$ {\bf all\,\, charged} $ &
 $122.7$ & $81.27$ & $60.10$ & $14.52$ & $4.39$ \\
 $\bf Energy$ &${\bf N_{coll}*E_{T}^{pA}}$ &
   $121.8$ & $84.33$ & $59.74$ & $11.50$ &  \\
  $ {\bf (GeV)}$   &${\bf N_{proj}*E_{T}^{pA}}$ &
  $80.90$ & $50.22$ & $36.17$ & $7.6$ &  \\
\hline
\hline
 ${\bf  Mean\,\, Transverse}$ &$ {\bf all\,\,neutrals}$ &
 $89.87$ & 59.43 & $44.10$ & $10.64$ & $3.15$ \\
 $\bf Energy$ &${\bf N_{coll}*E_{T}^{pA}}$ &
 $87.41$ & $60.1$ & $42.87$ & $8.25$ &  \\
  $ {\bf (GeV)}$   &${\bf N_{proj}*E_{T}^{pA}}$ &
  $58.05$ & $36.03$ & $25.95$ & $5.44$ & \\
\hline
\hline
\end{tabular}
\end{table}

\newpage
\begin{table}
\caption{Comparison of average multiplicities calculated in 
  C - Air collisions at different energies with the expectation in 
   two different superposition models. $N_{p}$ is the average 
   number of projectile nucleons taking part in the inelastic 
   C - Air collision.The energies given are the energy per nucleon.}
\label{Tab4}
\vskip 0.5cm
\begin{tabular}{||c||c||c|c|c|c|c||}  
 \hline \hline
    &   &  &  &   &    &  \\
 ${ \bf  ENERGY}$ & ${\bf N_p}$ &${\bf n_{\pi^+}^{C-Air}}$ &
    ${\bf  n_{\pi^+}^{p-Air}}$ &  ${\bf  n_{\pi^+}^{p-p}}$ & 
${ \bf N_p n_{\pi^+}^{p-p}}$ & ${\bf N_p n_{\pi^+}^{p-Air}}$ \\
 ${\bf TeV}$     &     &   &  &   &   &  \\
 \hline
 \hline
${\bf 1 }$ & $2.23$ & $8.20$  & $2.24$ & $4.89$ & $10.90$ & $4.99$ \\
\hline
 \hline
 ${\bf 10}$ & $2.40$ & $13.96$ & $3.77$ & $7.27$ & $17.44$ & $9.05$ \\
 \hline
 \hline
 ${\bf 100} $ & $2.66$ & $27.11$ & $6.70$ & $10.78$ & $28.67$& $17.82$ \\
  \hline
 \hline
 ${\bf 1000}$ & $2.93$& $47.33$& $11.58$ &$16.16$& $47.34$& $33.92$ \\
   \hline
 \hline
 ${\bf 10000}$ & $3.44$& $101.70$ &$22.65$&$27.00$&$92.88$& $77.92$ \\
   \hline
 \hline
\end{tabular}
\end{table}

\begin{table}
\caption{Comparison of $Z_{\pi} $ spectrum weighted 
  moments calculated in 
  C - Air collisions at different energies with the expectation in 
   two different superposition models. $N_{p}$ is the average 
   number of projectile nucleons taking part in the inelastic 
   C - Air collision.The energies given are the energy per nucleon.}
\label{Tab5}
\vskip 0.5cm
\begin{tabular}{||c||c||c|c|c|c|c||}  
 \hline \hline
    &   &  &  &   &    &  \\
 ${ \bf  ENERGY}$ & ${\bf N_p}$ &${\bf Z_{\pi^+}^{C-Air}}$ &
    ${\bf  Z_{\pi^+}^{p-Air}}$ &  ${\bf  Z_{\pi^+}^{p-p}}$ & 
${ \bf N_p Z_{\pi^+}^{p-p}}$ & ${\bf N_p Z_{\pi^+}^{p-Air}}$ \\
${\bf TeV}$     &     &   &  &   &   &  \\
 \hline
 \hline
${\bf 1 }$ & $2.23$ & $0.047$  & $0.010$ & $0.026$ & $0.057$ & $0.022$ \\
\hline
 \hline
 ${\bf 10}$ & $2.40$ & $0.047$ & $0.012$ & $0.024$ & $0.058$ & $0.029$ \\
 \hline
 \hline
 ${\bf 100} $ & $2.66$ & $0.049$ & $0.010$ & $0.023$ & $0.061$& $0.027$ \\
  \hline
 \hline
 ${\bf 1000}$ & $2.93$& $0.054$& $0.011$ &$0.028$& $0.082$& $0.032$ \\
 \hline
 \hline
 ${\bf 10000}$ & $3.44$& $0.062$ &$0.012$&$0.024$&$0.082$& $0.041$ \\
 \hline
 \hline
\end{tabular}
\end{table}

\newpage

\begin{table}
\caption{Inelasticities in p - Air collisions as calculated with 
   HIJING model}. 
\label{Tab6}
\vskip 0.5cm
\begin{tabular}{||c||c|c|c|c|c|c|c|c||}  
 \hline \hline
 ${ \bf  ENERGY}$ & ${\bf K_{\pi^+}}$ &${\bf K_{\pi^-}}$ &
    ${\bf  K_{\pi^0}}$ &  ${\bf  K_{K^+}}$ & 
 ${\bf K_{K^-}}$ &${\bf K_{ch}}$ &${\bf K_{neu}}$ &${\bf K_{all}}$ \\
${\bf TeV}$&   &   &  &   &   &  &  &   \\
 \hline
 \hline
${\bf 1 }$ & $0.053$ & $0.045$  & $0.052$ & $0.0076$ & $0.0070$ & 
 $0.227$& $0.138$ & $0.366$ \\
\hline
 \hline
 ${\bf 10}$ & $0.056$ & $0.047$ & $0.054$ & $0.0080$ & $0.0070$ & 
 $0.240$ & $0.150$ & $0.390$ \\
 \hline
 \hline
 ${\bf 100} $ & $0.060$ & $0.052$ & $0.060$ & $0.0092$ & $0.0080$& 
  $0.271$ & $0.165$ & $0.435$ \\
  \hline
 \hline
 ${\bf 1000}$ & $0.068$& $0.059$& $0.066$ &$0.0100$& $0.0100$& 
  $0.289$ & $0.183$ & $0.472$ \\
   \hline
 \hline
 ${\bf 10000}$ & $0.080$& $0.072$  & $0.082$  & $0.0126$  &  $0.0123$& 
 $0.306$  &$0.215$  & $0.522$   \\
   \hline
 \hline
\end{tabular}
\end{table}

\begin{table}
\caption{Comparison of $Z_{\pi} $ and $ Z_{K}$ moments calculated in 
  p - Air collisions at different energies with the expectation in 
    different  models : HIJING, DPMJET\protect{\cite{ranf94a}},
 \protect{\cite{ranf94b}}, HEMAS\protect{\cite{forti90}} 
and SIBYLL ,\protect{\cite{flet94}}.} 
\label{Tab7}
\vskip 0.5cm
\begin{tabular}{||c||c|c|c|c|c|c|c|c||}  
\hline \hline
    & {\bf HIJ}  &  DPM &  HEMAS &  SIBYLL  &
     HIJ &   DPM &  HEMAS &  SIBYLL  \\
    \hline \hline
 ${ \bf  ENERGY}$ & ${\bf Z_{\pi}}$ &${\bf Z_{\pi}}$ &
    ${\bf  Z_{\pi}}$ &  ${\bf  Z_{\pi}}$ & 
 ${\bf Z_K}$ &${\bf Z_K}$ &${\bf Z_K}$ &${\bf Z_K}$ \\
${\bf TeV}$&   &   &  &   &   &  &  &   \\
 \hline
 \hline
${\bf 1 }$ & $0.046$ & $0.067$  & $0.061$ & $0.072$ & $0.0080$ & 
 $0.0098$& $0.0104$ & $0.0073$ \\
\hline
 \hline
 ${\bf 10}$ & $0.043$ & $0.069$ & $0.057$ & $0.068$ & $0.0071$ & 
 $0.0099$ & $0.0113$ & $0.0071$ \\
 \hline
 \hline
 ${\bf 100} $ & $0.040$ & $0.068$ & $0.056$ & $0.067$ & $0.0066$& 
  $0.0102$ & $0.0116$ & $0.0070$ \\
  \hline
 \hline
 ${\bf 1000}$ & $0.040$& $0.066$& $0.056$ &$0.066$& $0.0068$& 
  $0.0101$ & $0.0123$ & $0.0070$ \\
   \hline
 \hline
 ${\bf 10000}$ & $0.042$&  &  &   &  $0.0076$& 
   &   &   \\
 \hline
 \hline
\end{tabular}
\end{table}

\newpage

%Figure 1

\begin{figure}[h]
\vspace{3cm}
\hspace{0.5cm}
\psfig{figure=fig01.epsi,height=4.5in,width=4.5in}
\vspace{1.cm}
\caption{Rapidity (Fig. 1a) and transverse momenta(Fig. 1b) 
distributions of $\Lambda^0$ (dotted histograms),
 $\bar{\Lambda^0}$ (dashed histograms),
and $K_{S}^{0}$ (full histograms) produced in $pp$ interactions 
at $300\,GeV$. HIJING  results are shown by histograms.
The experimental data are taken from  Lo 
  Pinto et al.\protect{\cite{ex80}}.}
\end{figure}

\newpage

%Figure 2
\begin{figure}[h]
\vspace{3cm}
\hspace{0.5cm}
\psfig{figure=fig02.epsi,height=4.5in,width=4.5in}
\vspace{1cm}
\caption{Rapidity distributions of  $\Lambda$ and 
  $\bar{\Lambda}$ produced  in  central
 SS collisions at 200A GeV  (Fig. 2a and Fig. 2b) respectively.
In parts (b)  the $p-\bar{p}$ distributions are
shown while parts  (d) correspond to the $\Lambda-\bar{\Lambda}$
distributions for S + Ag, S + Au interactions at 200A GeV. 
 Expectations based on HIJING
 model are depicted as solid  histograms .
 The theoretical predictions based on VENUS model 
 are depicted as dotted histograms (option without final 
 state interaction)and as dashed histograms (option including final state
interactions ) The NA35 data (full circles) are from 
 Alber et al. \protect{\cite{na35_5}} for part (a) and (b) and from 
 R\"orich et al. \protect{\cite{na35_7}} for part (c) and (d).
The open circles show the distributions 
 for SS collisions reflected  at midrapidity.}
\end{figure}

%Figure 3
\begin{figure}[h]
\vspace{3cm}
\hspace{0.5cm}
\psfig{figure=fig03.epsi,height=4.5in,width=4.5in}
\vspace{1cm}
\caption{Comparison of central $S+S$  collisions at 200 AGeV
(Fig. 3a -antiproton and Fig. 3b - negative kaons) 
with central $PbPb$ collisions at 160 AGeV
(Fig. 3c-antiproton and Fig. 3d- negative kaons).
The experimental data are from NA44\protect{\cite{murray96}} for 
antiprotons and from NA35\protect{\cite{marg96}} for negative kaons.
The open circles show the distributions
for SS collisions reflected at mid-rapidity. The solid,
dashed and dotted histograms have the same meaning as in Fig. 2.}
\end{figure}

%Figure 4
\begin{figure}[h]
\vspace{3cm}
\hspace{0.5cm}
\psfig{figure=fig04.epsi,height=4.5in,width=4.5in}
\vspace{1cm}
\caption{Comparison of central $S+S$ at 200 AGeV (a,b)
NA35\protect{\cite{na35_7,na35_4}} and central
$Pb+Pb$ at $160$ AGeV (c,d) (NA49\protect{\cite{na49_4,na49_1,na49_2}}
(solid circles),
NA44\protect{\cite{xu_96}}(solid squares)) data with calculations.
Open circles and open squares are reflected data around mid-rapidity.
Solid and dashed histograms correspond to
HIJING  and  VENUS  models respectively.}
\end{figure}

%Figure 5
\begin{figure}[h]
\vspace{3cm}
\hspace{0.5cm}
\psfig{figure=fig05.epsi,height=4.5in,width=4.5in}
\vspace{1cm}
\caption{Pseudorapidity distributions for transverse energy of 
secondary produced particles: all secondaries (Fig. 5a) ,
 all neutrals( Fig. 5b) , all charged (Fig. 5c)  and gluons
 (Fig. 5d) in  $p+Air$ Nucleus interactions .
The dotted(for 1 TeV laboratory energy), dashed(for 100 TeV 
laboratory energy) and solid (for 1000 TeV laboratory energy) 
histograms are theoretical values given by HIJING  model.}
\end{figure}

%Figure 6
\begin{figure}[h]
\vspace{3cm}
\hspace{0.5cm}
\psfig{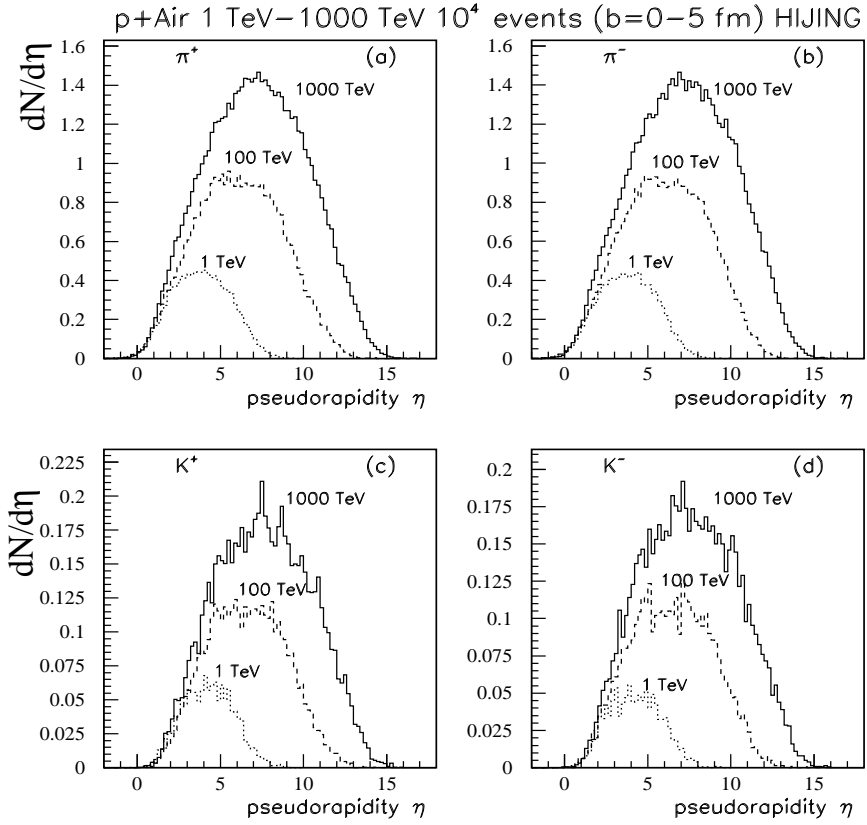}
\vspace{1cm}
\caption{Pseudorapidity distributions for 
 $p+Air \rightarrow \pi^+ + X $ (Fig. 6a),
 $p+Air \rightarrow \pi^-+ X $ (Fig. 6b),
 $p+Air \rightarrow K^+ + X $ (Fig. 6c),
 $p+Air \rightarrow K^- + X $ (Fig. 6d) , at 1 TeV, 100 TeV and 1000 TeV.
 The dotted, dashed and solid histograms have the same meaning as in 
 Figure 5.}
\end{figure}

%Figure 7
\begin{figure}[h]
\vspace{3cm}
\hspace{0.5cm}
\psfig{figure=fig07.epsi,height=4.5in,width=4.5in}
\vspace{1cm}
\caption{Pseudorapidity distributions for
 $p+Air \rightarrow \pi^o + X $ (Fig. 7a),
  $p+Air \rightarrow \gamma+ X $ (Fig. 7b),
 $p+Air \rightarrow \Lambda^0 + X $ (Fig. 7c),
  $p+Air \rightarrow D^{\pm} + X $ (Fig. 7d),at 
 1 TeV, 100 TeV and 1000 TeV.
   The dotted, dashed and solid histograms have 
   the same meaning as in Figure 5.}
\end{figure}

% Figure 8
\begin{figure}[h]
\vspace{3cm}
\hspace{0.5cm}
\psfig{figure=fig08.epsi,height=4.5in,width=4.5in}
\vspace{1cm}
\caption{Transverse momenta distributions for
 $p+Air \rightarrow all\,\, charged + X $ (Fig. 8a),
 $p+Air \rightarrow p +X $ (Fig. 8b),
 $p+Air \rightarrow \pi^{+}+ X $ (Fig. 8c),
 $p+Air \rightarrow \pi^{-} + X $ (Fig. 8d),
 at 1 TeV, 100 TeV and 1000 TeV.
 The dotted, dashed and solid histograms have 
 the same meaning as in Figure 5.}
\end{figure}

% Figure 9
\begin{figure}[h]
\vspace{3cm}
\hspace{0.5cm}
\psfig{figure=fig09.epsi,height=4.5in,width=4.5in}
\vspace{1cm}
\caption{Test of Feynman scaling in the production of 
 $p+Air\rightarrow \pi^{\pm}+X$ collisions 
 (Fig. 9a) and $p+Air\rightarrow p+\bar{p} + X$ collisions (Fig. 9b),
 between 1 TeV - 1000 TeV.
 The Feynman -$ x_F$ distributions were calculated with HIJING model.
  The dotted, dashed and solid histograms have
 the same meaning as in Figure 5.}
\end{figure}

% Figure 10
\begin{figure}[b]
\vspace{3cm}
\hspace{0.5cm}
\psfig{figure=fig10.epsi,height=4.5in,width=4.5in}
\vspace{1cm}
\caption{Charged multiplicities distributions in $p+Air$ 
interactions at 1000 TeV.Contributions from all events
are depicted in Fig. 10a.
In Fig. 10b the histogram  is  from HIJING model
calculations with contributions from events with $N_{jet}=0$,
Fig. 10c the histogram is  from calculations with
contributions from events with $N_{jet}=1$ and in 
Fig. 10d  the histogram is from calculations with
contributions from events with $N_{jet}>1$.}
\end{figure}

% Figure 11
\begin{figure}[h]
\vspace{3cm}
\hspace{0.5cm}
\psfig{figure=fig11.epsi,height=4.5in,width=4.5in}
\vspace{1cm}
\caption{Pseudorapidity distributions for all secondaries produced 
in Fe+Air interactions(Fig. 11a),S+Air interactions(Fig. 11b),
Ne+Air interactions(Fig. 11c)and He+Air interactions (Fig. 11d)
at 17.86 TeV/Nucleon for events generated in
different impact parameters intervals ($b_{min},b_{max}$).
See the text for explanations.}
\end{figure}

%Figure 12
\begin{figure}[h]
\vspace{3cm}
\hspace{0.5cm}
\psfig{figure=fig12.epsi,height=4.5in,width=4.5in}
\vspace{1cm}
\caption{Pseudorapidity distributions for transverse energy of
secondary produced particles :all secondaries (Fig. 12a) ;
all neutrals (Fig. 12b) ; all charged (Fig. 12c) ; gluons (Fig. 12d)
in $A+Air$ interactions at 17.86 TeV/Nucleon.
The theoretical values were calculated with HIJING model
and are depicted by dotted (He+Air) ; dot-dashed (Ne+Air) ;
dashed (S+Air) and solid (Fe+Air) histograms.}
\end{figure}

% Figure 13
\begin{figure}[b]
\vspace{3cm}
\hspace{0.5cm}
\psfig{figure=fig13.epsi,height=4.5in,width=4.5in}
\vspace{1cm}
\caption{Pseudorapidity distributions for transverse momenta of
secondary particles for $A+Air\rightarrow \pi^{\pm}+X$ (Fig. 13a)
and $A+Air \rightarrow p+X$ collisions (Fig. 13b)
at 17.86 TeV/Nucleon.
The histograms have the same meaning as in Figure 12.}
\end{figure}

% Figure 14
\begin{figure}[b]
\vspace{3cm}
\hspace{0.5cm}
\psfig{figure=fig14.epsi,height=4.5in,width=4.5in}
\vspace{1cm}
\caption{Pseudorapidity distributions for
$A+Air \rightarrow \pi^+ + X $ (Fig. 14a),
$A+Air \rightarrow \pi^-+ X $ (Fig. 14b),
$A+Air \rightarrow K^+ + X $ (Fig. 14c),
$A+Air \rightarrow K^- + X $ (Fig. 14d), at 17.86 TeV/Nucleon.
The histograms have the same meaning as in Figure 12.}
\end{figure}

% Figure 15
\begin{figure}[h]
\vspace{3cm}
\hspace{0.5cm}
\psfig{figure=fig15.epsi,height=4.5in,width=4.5in}
\vspace{1cm}
\caption{Transverse momenta distributions for
$A+Air \rightarrow all\,\, charged + X $ (Fig. 15a),
$A+Air \rightarrow p +X $ (Fig. 15b),
$A+Air \rightarrow \pi^++ X $ (Fig. 15c),
$A+Air \rightarrow \pi^- + X $ (Fig. 15d),at 17.86 TeV/Nucleon.
The histograms have the same meaning as in Figure 12.}
\end{figure}

% Figure 16
\begin{figure}[h]
\vspace{3cm}
\hspace{0.5cm}
\psfig{figure=fig16.epsi,height=4.5in,width=4.5in}
\vspace{1cm}
\caption{The Feynman $x_{F}$ distributions in the production of
$A+Air \rightarrow \pi^{\pm}+X$ collisions (Fig. 16a) and
$A+Air \rightarrow p+\bar{p}+X$ collisions(Fig. 16b), at 17.86
TeV/Nucleon.
The histograms have the same meaning as in Figure 12 .}
\end{figure}

% Figure 17
\begin{figure}[b]
\vspace{3cm}
\hspace{0.5cm}
\psfig{figure=fig17.epsi,height=4.5in,width=4.5in}
\vspace{1cm}
\caption{Charged particles multiplicities distributions in
Fe+Air interactions (Fig. 17a) and S+Air interactions 
(Fig. 17b) at 17.86 TeV/Nucleon.
Contributions from events with number of minijets
$N_{jet} > 1$ are represented in Fig. 17c for Fe+Air interactions
and in Fig. 17d for S+Air interactions.}
\end{figure}

% Figure 20 air03
% Figure 18 air0f
\begin{figure}[h]
\vspace{3cm}
\hspace{0.5cm}
\psfig{figure=fig18.epsi,height=4.5in,width=4.5in}
\vspace{1cm}
\caption{Comparison of Feynman - $x_{F}$ distributions 
of $\pi^{-}$ mesons produced in proton - proton collisions 
at 250 GeV ( Fig. 18a).The  experimental data are from the 
EHS - NA22 Collaboration\protect{\cite{adam88}} ; 
for proton - proton collisions at 360 GeV
(Fig. 18b) .The experimental data are from 
Bailly et al.\protect{\cite{bail87,bail86}}.The histogram are 
theoretical predictions for 360 GeV (dashed) and for 400 GeV (solid) .
Comparison of Feynman - $x_{F}$ distributions 
of $K^{+}$ - mesons produced in 
proton - proton collision at 400 GeV (Fig. 18c).The experimental data 
are from LEBC - EHS Collaboration\protect{\cite{agui91}} ;
Comparison of Feynman - $x_{F}$ distributions of $\pi^{+}$ -
 mesons produced in proton - proton collision at 205 GeV 
(Fig. 18d). The experimental data 
are from Kafka et al.\protect{\cite{kaf77}}.}
\end{figure}

% Figure 22 air03 
% Figure 19 air0f
\begin{figure}[h]
\vspace{3cm}
\hspace{0.5cm}
\psfig{figure=fig19.epsi,height=4.5in,width=4.5in}
\vspace{1cm}
\caption{Comparison of Feynman $x_{F}$ distributions of 
 $\Lambda^{0}$ (Fig. 19a) and 
 $K_{s}^{0}$ (Fig. 19b) produced in proton - proton collisions at 360 GeV .
The histogram are theoretical predictions for 
360 GeV (dashed) and for  400 GeV (solid). The experimental data 
are from Bailly et al.\protect{\cite{bail87}}.
Feynman $x_{F}$ distributions of $\Lambda^{0}$ (Fig. 19c) and 
$ \gamma $ (Fig. 19d) in energy region 1 TeV - 1000 TeV.
The histograms are HIJING model predictions for 1 TeV (solid),
10 TeV (dashed),100 TeV (dotted) and 1000 TeV (dot dashed).}
\end{figure}

% Figure 23 air03
%Figure 20 air0f
\begin{figure}[b]
\vspace{3cm}
\hspace{0.5cm}
\psfig{figure=fig20.epsi,height=4.5in,width=4.5in}
\vspace{1cm}
\caption{Comparison of Feynman $x_{F}$ distributions 
of $\pi^{-}$ - mesons produced in $\pi^{+}+ p$ collisions at 
250 GeV (Fig. 20a). The  experimental data are from the 
EHS - NA22 Collaboration 
\protect{\cite{adam88}} ; $\pi^{+}$ - mesons produced in 
$\pi^{+}+ p$ collisions at 250 GeV (Fig. 20b). 
The experimental data are from
Brenner et al. \protect{\cite{bren82}}.
Comparison of Feynman $x_{F}$ distributions of positives charged $h^{+}$
produced in $\pi^{+}+ p$ collisions at 250 GeV (Fig. 20c).
The experimental data are from the EHS - NA22 Collaboration 
\protect{\cite{adam88}}.
Comparison of Feynman $x_{F}$ distributions of $\pi^{-}$ - mesons
produced in $\pi^{+}+ p$ collisions at 175 GeV (Fig. 20d) .
The experimental data are from Brenner et al.\protect{\cite{bren82}}.}
\end{figure}

\end{document}